\documentclass[twocolumn,showpacs,preprintnumbers,amsmath,amssymb]{revtex4}
\usepackage{bm}% bold math

\newcommand{\be}{\begin{equation}}
\newcommand{\ee}{\end{equation}}
\newcommand{\bea}{\begin{eqnarray}}
\newcommand{\eea}{\end{eqnarray}}
\newcommand{\nn}{\nonumber}
\newcommand{\p}{\phi}
\newcommand{\pp}{\tilde \phi}
\newcommand {\pb}{\bar \phi}

\begin{document}
\title{Observing an open FRW de Sitter universe living in a Minkowski spacetime.}
\author{F. Loran}
\email{loran@cc.iut.ac.ir}

 \affiliation{Department of  Physics, Isfahan University of Technology
 (IUT), Isfahan,  Iran}

\begin{abstract}
 We show that people living in a four dimensional Minkowski spacetime and located in the Fubini vacua of an
 unstable critical scalar theory, observe an open FRW de~Sitter universe.
\end{abstract}

 \maketitle

\section{Introduction}
 The WMAP results \cite{WMAP} combined with earlier cosmological observations
 shows that we are living in an accelerating universe. The currently observed lumpiness in the temperature
 of the cosmic microwave background is just right for a flat
 universe though there are also some evidences that our universe
 is spatially open \cite{Gott}. The great simplifying fact of cosmology is that the universe appears to be homogeneous
 and isotropic along a preferred set of spatial hypersurfaces \cite{16}. Of course
 homogeneity and isotropy are only approximate, but they become
 increasingly good approximations on larger length scales, allowing
 us to describe spacetime on cosmological scales by the
 Robertson-Walker metric.
 Constructing four dimensional de Sitter vacuum as a string theory
 (M-theory) solution has been a long standing challenge.
 An outstanding example of string theory models of de~Sitter vacua are the
 KKLT models \cite{KKLT} with an exponentially
 large number of stable and metastable vacua without supersymmetry
 or with ${\cal N}=1$ supersymmetry in four dimensions, the "landscape" \cite{Sus}.  In KKLT
 models,  metastable de Sitter vacua of type
 IIB string theory is constructed by adding $\overline{\mbox{D3}}$-branes  to the
 GKP \cite{GKP} model of highly warped IIB compactifications with
 nontrivial NS and RR three-form fluxes after certain fine tuning of
 the fluxes.

 Recently, we realized that it is  possible to observe a de~Sitter universe while living in a flat
 background.  This proposal was the consequence of a simple observation: the fluctuations of the scalar field around the
 classical trajectory of an unstable massless $\p^4$ model in four dimensional flat
 Euclidean spacetime is governed by a conformally  coupled scalar field theory in four dimensional
 de Sitter background \cite{Solitons}. This classical trajectory is
 the Fubini vacua of the classically conformal-invariant scalar
 field  theories. In \cite{Fubini} S. Fubini verified that critical
 scalar theories possess a classical vacua with $O(D,1)$ symmetry in which the expectation
 value of the scalar field is non-vanishing. The motivation to study such a classical vacua at that
 time was "to introduce a fundamental scale of hadron phenomena by
 means of dilatation non-invariant vacuum state in the frame work of
 a scale invariant Lagrangian field theory" \cite{Fubini}.
 This result is interesting due to its uniqueness. In four dimensions, in principle,
 one can consider two classes of critical (classically scale-free ) scalar field
 theories i.e. massless $\phi^4$ models on Euclidean spacetime
 with $g$, the coupling constant, either positive or negative (we assume the potential
 $V(\p)=-\frac{g}{4}\p^4$). Although scalar theory with $g>0$ seems
 to be not physical as the potential is not bounded from below but in this case,
 the Euler-Lagrange equation of motion has an interesting
 classical solution say $\p_0$ with finite action $S[\p_0]\sim
 g^{-1}$. For $g<0$, one can still consider a solution like
 $\p_0$ obtained by an analytic continuation from $g>0$ to
 $g<0$ region. But such a solution is singular on the surface of a
 sphere which radius is proportional to $g$. Consequently the action $S[\p_0]$ is infinite and $\p_0$ can not
 be considered as a classical trajectory. For $g>0$ it is shown in \cite{Solitons} that the information geometry of the
 moduli space of $\p_0$ given by Hitchin formula \cite{Hit} is  Euclidean $\mbox{AdS}_5$,
 \be
 {\cal G}_{IJ}d\theta^Id\theta^J=\frac{1}{\beta^2}\left(d\beta^2+d
 a^2\right),
 \label{Mod-met}
 \ee
 where $\theta^I\in\{\beta,a^\mu\}$ and $I=1,\cdots,5$. The moduli here are $a_\mu$'s the location of the center
 of $\p_0$ and $\beta$ which is proportional to the inverse of the size of $\p_0$.   This resembles the information
 geometry of SU(2) instantons. In addition $V(\p_0)$ can be shown to be proportional
 to the SU(2) one-instanton density \cite{Blau}. Interestingly,
 $\p_0$ is the bulk to boundary propagator in the $\mbox{AdS}_5$ geometry
 of the moduli space.

 In ref. \cite{U1} we generalized the model to scalar theory
 coupled to U(1) gauge field. Such a generalization is essential as it shows how
 by optical observations people living in a flat Euclidean space observe
 a de~Sitter geometry for their universe. In this model the massless scalar field is
 charged though we have not observed light charged scalars. This
 problem can be resolved noting that as we will show, in this model observations are made in a de~Sitter background
 in which the scalar field appears to be conformally coupled to the de~Sitter background. Therefore its mass is
 proportional to the scalar curvature $R$ of the observed
 universe. Using the WKB approximation, the lifetime of the observed de~Sitter background is
 calculated in \cite{U1} and is shown to be proportional to $e^{g^{-1}}$. (To my knowledge, this
 result is given for the first time in a beautiful paper by Coleman where $\p_0$ is called a
 "bounce" \cite{Coleman}.)
 Consequently in the weak coupling limit $g\to{^+0}$ the lifetime increases exponentially.

 In this paper we study the critical scalar theory on Minkowski spacetime. Here $\p_0$ is singular on a hyperbola in
 the timelike region which asymptotes to the lightcone. The total energy
 of $\p_0$, measured by an observer located at the center of $\p_0$ is conserved and
 vanishing though the energy density is a function of space and time. The energy density diverges
 in the neighborhood of singularity causing a gravitational collapse when the scalar theory is coupled to gravity.
 Fortunately the singularity is safe. On the one hand in Minkowski spacetime, the distance between the observers and the
 singularity is proportional to $\beta$. On the other hand there
 is some mechanism of $\beta$ transition in the model: larger $\p_0$'s decay to smaller ones due to say thermal
 fluctuations
 around $\p_0$ and finally there remains only a gas of zero sized bubbles which corresponds to $\beta\to\infty$.
 The mechanism of such transition is not clear yet but its phenomenology, probably is similar
 to that of the discretuum of possible de Sitter vacua in KKLT models
 \cite{fall}. Therefore, for the most stable $\p_0$ solution, the singularity is located at infinite
 future and is out of reach. Furthermore, from the observers point of view, the
 observable universe is an open de~Sitter space which horizon is located on the
 singularity. Therefore they do not see the singularity at~all for
 any value of $\beta$! Of course they should feel some back reactions when the scalar theory is coupled to gravity
 caused by the $\beta$ transition.  The energy density $\rho$ and pressure $p$ that they measure are
 constants satisfying the dark energy equation of state $\rho=-p=\Lambda$
 ($8\pi G=1$), where $\Lambda$ is the cosmological constant. A question here is the value of $\Lambda$ or equivalently
 $R$, the curvature scalar. At classical level, $R$ is not determined in the critical scalar model as is expected.
 Because the theory is
 classically scale free. The quantum theory is not scale free due to loop corrections. Therefore quantum
 corrections are the hopeful candidates to give the value of the observed scalar curvature. The details are not
 clear for us yet and we postpone it to future works.

 In Minkowski spacetime in the case of critical scalar model with negative coupling constant the
 singularity of $\p_0$ is a hyperbola in the spacelike region which asymptotes  to the
 lightcone. One can easily verify that in this case the total
 energy for existence of $\p_0$ is infinite. Thus, similar to
 the Euclidean case, one can conclude that $\p_0$ uniquely exists only in
 the unstable ($g>0$) critical scalar model.

 The organization of the paper is as follows. In the next section
 we study the critical scalar theory on flat Euclidean background and
 determine the role of the moduli $\beta$ in the stability of the
 solutions. We show that by recasting the scalar theory in terms of new fields $\pp=\p-\p_0$ at the end of the day
 one obtains a $\p^4$ model conformally coupled to a de~Sitter background.
 In section \ref{M} we switch to the Minkowski spacetime by a
 Wick rotation $t\to i t$ and study the observed de~Sitter universe in terms of
 the Robertson-Walker metric.
 %-----------------------------------------------------------------------------------------
 \section{Critical scalar theory in $D=4$ Euclidean space}\label{Cr}
 In this section we study scalar field theories in four dimensional Euclidean space
 invariant under rescaling transformation $x\to x'=\lambda x$,
 $\lambda>0$. There are in general three scale-free scalar theories: $\p^4$
 model in $D=4$ and  $\p^3$ and $\p^6$ models in $D=6,3$
 respectively. In this paper we only study $\p^4$ model in $D=4$
 though the main result of this paper can be simply generalized to the
 other two scalar models.  The action in Euclidean space is
 \be
 S[\p]=\int d^4x
 \left(\frac{1}{2}\delta^{\mu\nu}\partial_\mu\p\partial_\nu\p-
 \frac{g}{4}\p^4\right)
 \label{action}
 \ee
 where we assume $g>0$. Consequently the potential $V(p)\sim
 -\p^4$ and is not bounded from below. The Kronecker delta symbol $\delta^{\mu\nu}$ stands for the metric of flat
  Euclidean space.  The corresponding equation of
 motion is a non-linear Laplace equation $\nabla^2 \p+g\p^3=0$,
 where $\nabla^2=\delta^{\mu\nu}\partial_\mu\partial_\nu$. One can easily
 show that for $g>0$, a solution of the non-linear Laplace equation is
 \be
 \p_0(x;\beta,a^\mu)=\sqrt{\frac{8}{g}}
 \frac{\beta}{\beta^2+(x-a)^2},
 \label{soliton}
 \ee
 where $(x-a)^2=\delta_{\mu\nu} (x-a)^\mu(x-a)^\nu.$ $\beta$ and $a^\mu$ are
 undetermined parameters describing the
 the size and location of $\p_0$. These moduli are consequences
 of symmetries of the action i.e. invariance under rescaling and
 translation. The information geometry of the moduli space, given by
 Hitchin formula \cite{Hit}
 \be
 {\cal G}_{IJ}=\frac{1}{N}\int d^4x
 {\cal{L}}_0\partial_I\left(\log{\cal{L}}_0\right)
 \partial_J\left(\log{\cal{L}}_0\right),
 \label{InG}
 \ee
 is an Euclidean $\mbox{AdS}_5$ space (\ref{Mod-met}).
 $N=\frac{4^3}{5}\int d^4 x {\cal L}_0$ is
 a normalization constant and ${\cal{L}}_0=\frac{g}{4}\p_0^4$ is the Lagrangian density
 calculated at $\p=\p_0$.
 The moduli $a^\mu$ are present since the action is invariant under
 translation. The existence of $\beta$ is the result of invariance under
 rescaling \cite{U1}.

 To my knowledge, the solution $\p_0$ is obtained for the first time by Fubini.
 He looked for a solution of the equation of motion "in which the
 vacuum expectation value of the field $\p(x)$ is non-vanishing" \cite{Fubini}.
 He verified that this vacua is not invariant under
 the Poincare group but is invariant under the de~Sitter group
 $O(3,1)$. Consequently by recasting the action in terms of new fields $\pp=\p-\p_0$
 one expects to obtain, after some field redefinitions, a scalar theory in de~Sitter
 background. In fact the action in terms of $\pp$  is,
 \be
 S[\p]=S[\p_0]+S_{\mbox{free}}[\pp]+S_{\mbox{int}}[\pp],
 \label{act}
 \ee
 where $S[\p_0]=\int d^4x {\cal L}_0=\frac{8\pi^2}{3g}$, and
 \be
 S_{\mbox{free}}[\pp]=\int d^4 x \left(
 \frac{1}{2}\delta^{\mu\nu}\partial_\mu\pp\partial_\nu\pp+\frac{1}{2}M^2(x)\pp^2\right)
 \label{free}
 \ee
 in which,
 \be
 M^2(x)=-3g \p_0^2
 =-24\frac{\beta^2}{\left(\beta^2+(x-a)^2\right)^2}.
 \label{mass}
 \ee
 These equations show that $\p_0$ is a metastable local minima of the
 action. This can also be verified explicitly by numerical analysis of action (\ref{action}), see ref. \cite{U1}.
 Equation (\ref{free}) can be used to show that the
 stability increases as $\beta\to \infty$. In fact if we calculate the variation of action
 at the stationary point $\p_0(\beta)$ for different values of the moduli $\beta_1$ and $\beta_2$,
 under variation $\delta\p$, from
 Eqs.(\ref{free},\ref{mass}) one verifies that,
 \bea
 \Delta S&=&\delta S|_{\beta_1}-\delta S|_{\beta_2}\nn\\
 &\sim& \int d^4x
 \left(\p_0(\beta_2)^2-\p_0(\beta_1)^2\right)\delta\p^2+{\cal
 O}(\delta\p^3).
 \eea
 For simplicity we assume that $a^\mu_i=0$, $i=1,2$.
 Therefore $\Delta S$ is proportional to,
 \be
 (\beta_1^2-\beta_2^2)\int_0^\infty dx
 \frac{x^3(-x^4+\beta_1^2\beta_2^2)}{(\beta_1^2+x^2)^2(\beta_1^2+x^2)^2}\delta\p^2.
 \label{stability}
 \ee
 For $\delta\p$ with compact support, i.e. $\delta\p=0$ {\em
 if} $\left|x\right|>\sqrt{\beta_1\beta_2}$ the integral above is positive therefore
 $\Delta S\sim(\beta_1^2-\beta_2^2)$. As far as $\p_0$ is a
 metastable local minima there exist $\delta \p$ with compact
 support such that $\delta S|_{\beta_i}>0$ $i=1,2$. Consequently if
 $\beta_1>\beta_2$ then $\delta S|_{\beta_1}>\delta
 S|_{\beta_2}>0$. One can convince herself/himself that for
 some $\delta\p$ one obtains $\delta S|_{\beta_2}<0$
 while $\delta S|_{\beta_1}>0$. Consequently one concludes that
 there is a transition  $\beta_2\to\beta_1$ induced by say,
 thermal fluctuations. In addition the stability increases as $\beta\to\infty$.

  The mass term in Eq.(\ref{free}) can be interpreted as interaction with the  background $\p_0$.
 Now recall that in general, by inserting $\pp=\sqrt{\Omega}\pb$ and
 $\delta_{\mu\nu}=\Omega^{-1}g_{\mu\nu}$ in the action
 $S[\pp]=\int d^4x \frac{1}{2}\delta^{\mu\nu}\partial_\mu\pp\partial_\nu\pp$,
 one obtains,
 \be
 S[\pp]=\int d^4x \sqrt{g}\left(\frac{1}{2}g^{\mu\nu}\partial_\mu\pb\partial_\nu\pb+
 \frac{1}{2}\xi R\pb^2
 \right),
 \ee
 i.e. a scalar theory on conformally flat background given by the metric
 $g_{\mu\nu}=\Omega\delta_{\mu\nu}$ in which  $\Omega>0$ is an arbitrary
 ${\cal C}^\infty$ function. $R$ is the scalar curvature of the
 background and $\xi=\frac{1}{6}$ is the conformal coupling
 constant. For details see \cite{Ted} or appendix C of
 \cite{Solitons}.
 Thus, defining $\pb=\Omega^{\frac{-1}{2}}\pp$,
 one can show that $S_{\mbox{free}}[\pp]$ given in Eq.(\ref{free})
 is the action of the scalar
 field $\pb$ on some conformally flat background,
 \be
 S_{\mbox{free}}[\pp]=\int d^4 x \sqrt{\left|g\right|}
 \left(\frac{1}{2}g^{\mu\nu}\partial_\mu\pb
 \partial_\nu\pb+\frac{1}{2}(\xi R+m^2)\pb^2\right).
 \label{Curvedaction}
 \ee
 with metric
 \be
 g_{\mu\nu}=\Omega\delta_{\mu\nu},\hspace{1cm}\Omega=\frac{M^2(x)}{m^2},
 \label{metric1}
 \ee
 where $m^2$ is the mass of $\pb$
 (undetermined) and $M^2(x)$ is given in Eq.(\ref{mass}).
 This result is surprising as one can show that the Ricci tensor
 $R_{\mu\nu}=\Lambda g_{\mu\nu}$, where $\Lambda=-\frac{m^2}{2}>0$ as far as $\Omega>0$.
 Consequently $\pb$ lives in a four dimensional
  de~Sitter space which scalar curvature $R=-2m^2$.
  The interacting part of the action,
  $S_{\mbox{int}}[\pp]=\int d^4x\sqrt{\left|g_{\mu\nu}\right|}{\cal L}_{\mbox{int}}$
  is well-defined in terms of $\pb$ on the corresponding $\mbox{dS}_4$:
  \be
  {\cal L}_{\mbox{int}}=
  -g\sqrt{\frac{-m^2}{3g}}\pb^3-\frac{g}{4}\pb^4.
  \ee
  Interestingly after a shift  of the scalar field $\pb\to\pb-\sqrt{\frac{-m^2}{3g}}$
  the action  (\ref{act}) can be written in the $\mbox{dS}_4$ as follows:
  \be
  S[\pb]=\int d^4 x \sqrt{\left|g\right|}
 \left(\frac{1}{2}g^{\mu\nu}\partial_\mu\pb
 \partial_\nu\pb+\frac{1}{2}(\xi
 R)\pb^2-\frac{g}{4}\pb^4\right).
  \label{R}
  \ee
  This is a scalar theory in a de~Sitter background with reversed
  Mexican hat potential. In a similar way, by recasting the critical scalar theory
  minimally  coupled to $U(1)$ gauge field in terms of
  fluctuations around the classical solution $\p=\p_0$ and
  $A_\mu=0$, one verifies that the action
  \be
  S=\int d^4
  x\left(\left|D_\mu\p\right|^2-\frac{g}{2}\left|\p\right|^4\right)+S_A,
  \label{Uact}
  \ee
 is equivalent to
 \be
 S=S[\p_0]+S[\pb,A_\mu]+S_A,
 \ee
 where
 \bea
 S[\pb,A_\mu]=
 \int d^4 x\sqrt{g}\left(\frac{1}{2}g^{\mu\nu}D_\mu\pb
 D_\nu{\pb}^*+V(\pb)\right),
 \eea
 $V(\pb)=\frac{1}{2}\xi
 R\left|\pb\right|^2-\frac{g}{4}\left|\pb\right|^4$
 and $S_A$ is the Kinetic term for the gauge field,
 \bea
 S_A&=&-\frac{1}{4}\int d^4x F_{\mu\nu}F_{\rho\sigma}\delta^{\rho\mu}\delta^{\sigma\nu}\nn\\
 &=&-\frac{1}{4}\int d^4 x
 \sqrt{g}g^{\mu\rho}g^{\nu\sigma}F_{\mu\nu}F_{\rho\sigma}.
 \eea
 $F_{\mu\nu}$ in the first equality above is the field strength in Minkowski
 spacetime. In the second equality $F_{\mu\nu}$ should be
 understood as the field strength on the de~Sitter space \cite{U1}.
 It should be noted that under the conformal transformation
 $g_{\mu\nu}\to\Omega g_{\mu\nu}$, in four dimensions $A_{\mu}\to
 A_\mu$.
 \section{The critical scalar theory in Minkowski spacetime}\label{M}
 The critical scalar theory in four dimensional Minkowski
 spacetime is given by the action
 \be
 S[\p]=\int d^4x
 \left(\frac{1}{2}\eta^{\mu\nu}\partial_\mu\p\partial_\nu\p+\frac{g}{4}\p^4\right),
 \ee
 where $\eta_{\mu\nu}=(+,-,-,-)$ and $g>0$. The equation of motion is a
 non-linear wave equation
 $\eta^\mu\partial_\mu\partial_\nu\p-g\p^3=0$ which has the
 solution
 \be
 \p_0=\sqrt{\frac{8}{g}}\frac{\beta}{\beta^2-(t-a^0)^2+\left|{\vec x}-{\vec a}\right|^2},
 \ee
 where ${\vec x }\in {\bf R}^3$.
 Here on we assume $a^\mu=0$ for simplicity. $\p_0$ is singular
 on the hyperbola $t^2=x^2+\beta^2$ and we define its distance to
 an observer located on the origin to be given by $\beta$. The
 Hamiltonian density ${\cal H}$ corresponding to $\p_0$, is,
 \be
 {\cal H}=\frac{16\beta^2}{g} \frac{t^2+x^2-\beta^2}{(-t^2+x^2+\beta^2)^4}
 \ee
 which tends to infinity in the vicinity of the singularity. As is explained in the introduction, using the results
 of section \ref{Cr} and the arguments after Eq.(\ref{stability}), we now that the most stable $\p_0$ is the
 zero-sized one,
 corresponding to $\beta\to\infty$. Therefore the singularity is safe when the scalar theory is coupled to gravity.
 For $t<\beta$ one can calculate, say, the total vacuum energy $H=\int d^3x {\cal H}$ corresponding to
 $\p_0$ which is surprisingly vanishing, $H=0$. Repeating the
 calculations of section \ref{Cr} one verifies that observers located
 at the origin of the Minkowski spacetime observe a de~Sitter
 space given by the conformally flat metric,
 \be
 ds^2=\frac{12 \beta^2}{\Lambda}\frac{1}{(\beta^2-t^2+x^2)^2}(-dt^2+d{\vec
 x}^2),
 \ee
 where $\Lambda>0$ is the cosmological constant. This metric can be obtained using Eq.(\ref{metric1})
 after a Wick rotation $t\to it$.
 We use a different set of coordinates in order to describe the
 observed de~Sitter space with FRW metric to see whether it is open,
 closed or flat. Defining, coordinates $u$, $\rho$ and $z_i$,
 $i=1, 2,3$ by the relations $z_i^2=1$, $t=u\cosh\rho$ and
 $x_i=u\sinh \rho z_i$ useful to describe the timelike region
 $t>\left|{\vec x}\right|$, one obtains,
 \be
 ds^2=\frac{12 \beta^2}{\Lambda}\frac{1}{(\beta^2-u^2)^2}\left(-du^2+u^2(d\rho^2+\sinh
 \rho^2 dz_i^2)\right).
 \ee
 we define a time coordinate $\tau$ by the relation
 $d\tau=\left(\beta^2-u^2\right)^{-1} du$. Thus one obtains,
 \be
 \tau=\left\{\begin{array}{llll}
 \frac{1}{\beta}\coth^{-1}\frac{u}{\beta}&&&u>\beta,\\
 \frac{1}{\beta}\tanh^{-1}\frac{u}{\beta}&&&u<\beta,
 \end{array}\right.
 \ee
 and
 \be
 ds^2=\frac{12\beta^2}{\Lambda}\left(-d\tau^2+\frac{\sinh^2(2\beta
 \tau)}{4\beta^2}(d\rho^2+\sinh^2\rho dz_i^2)\right)
 \ee
 One can call the region $u<\beta$ which can be observed by observers located on the
 origin the south pole and the $u>\beta$ region the north
 pole, a known terminology in de~Sitter geometry. The south pole and north
 pole in our model are separated by the horizon located at $u=\beta$, i.e the singularity of $\p_0$.
 By normalizing $\tau$ by the normalization factor
 $\sqrt{\frac{12}{\Lambda}}\beta$ and defining a new coordinate
 $r=\sinh\rho$, one at the end of the day obtains,
 \be
 ds^2=-d\tau^2+a(\tau)^2\left(\frac{dr^2}{1+r^2}+r^2dz_i^2\right),
 \ee
 in which
 $a(\tau)=\sqrt{\frac{3}{\Lambda}}\sinh\sqrt{\frac{\Lambda}{3}}\tau$.
 This is the Robertson-Walker metric for open de~Sitter universe.
 One can easily calculate the energy density $\rho$ and the pressure $p$ of the cosmological stuff
 corresponding to $\p_0$ using
 the  Friedmann equations for the open universe,
 \bea
 \left(\frac{\dot a}{a}\right)^2&=&\frac{8\pi
 G}{3}\rho+\frac{1}{a^2},\nn\\
 \frac{\ddot a}{a}&=&-\frac{4\pi
 G}{3}(\rho+3p).
 \eea
 One verifies that $p$ and $\rho$ satisfy the equation of state
 for the cosmological constant $\rho=-p=\Lambda$ ($8\pi G=1$).

 \section*{Acknowledgement}
 The author gratefully thanks A. Kusenko  and V. A. Rubakov for useful
 discussions.
 I also thank ICTP for hospitality while this work was completed.
 The financial support of Isfahan University of Technology (IUT) is
 acknowledged.
 %\newpage
 
\end{document}